# Coordinator Location Effects in AODV Routing Protocol in ZigBee Mesh Network

Abla Hussein
Department of Computer Science
Zarqa University
Zarqa, Jordan

Ghassan Samara
Internet Technology Department, Faculty of
Information Technology
Zarqa University, Zarqa, Jordan

## ABSTRACT
ZigBee mesh network is very important research field in computer networks. However, the location of ZigBee coordinator plays a significant role in design and routing performance. In this paper, an extensive study on the factors that influence the performance of AODV routing protocol had been performed through the study of battery voltage decaying of nodes, neighboring tables, time delay and network topology structure. Simulation results reveal that the location of the coordinator within approximate equal distances to all nodes is more appropriate for lifelong batteries and AODV routing performance.

## 1. INTRODUCTION
ZigBee is one of the newest technologies developed by ZigBee Alliance where its characteristics depend mainly on IEEE 802.15.4 standards which consist of both Medium Access Control (MAC) layer and Physical layers (PHY). ZigBee protocols have been designed to provide data communication through RF environments that are used commercially and in industry applications, this communication protocol stack is used for WSN applications [1].

ZigBee technology had been used widely in different commercial, medical, industrial and home automation applications, and the importance of keeping the network operating a longer time was the main objective of ZigBee manufacturers [2].

ZigBee devices had been designed to operate from several months to several years at low bit rate transmission in wireless networks applications at UHF and Microwave frequencies. ZigBee devices were classifies in three different categories, a coordinator that initiate the transmission, routers that build up routes between nodes and an end units that considered as final destination. ZigBee was manufactured according to specific application where each network required at least one coordinator to initiate the network operation in which a personal area network identifier (PAN ID) will be assigned and routes will be established to communicate among the different nodes of the network. Therefore, coordinators are the most important device in ZigBee network which is able to store information about node and mange network routes [3, 5]. Most of the research work in this field was more on mesh network topology and less on the impact of coordinator location on routing process.

In this research work, two different networks were analyzed, the first where the coordinator was located at the center of the network, and the second where the coordinator was located at the corner of the network and each network had two different scenarios of route connection, the 1st and the 2nd shortest routes. The work depends mainly on Ad-hoc On Demand Distance Vector Algorithm (AODV) as a routing protocol that communicates between routers and broadcast packets to destination for the two cases of Mesh Networks. The established neighboring table will support AODV routing protocol to direct packets toward destination in order to establish a shortest route, Figure .1 shows a flowchart of the routing process.

Fixed positioning of coordinator in different spots of the network will be researched and analyzed where routes schematics, voltage decline curves, neighboring tables and energy maps were the concluded results.

The rest of this paper is organized as following: In section 2 related works, the simulation steps present in Section 3. simulation analysis in section 4 and section 5 concludes the results.

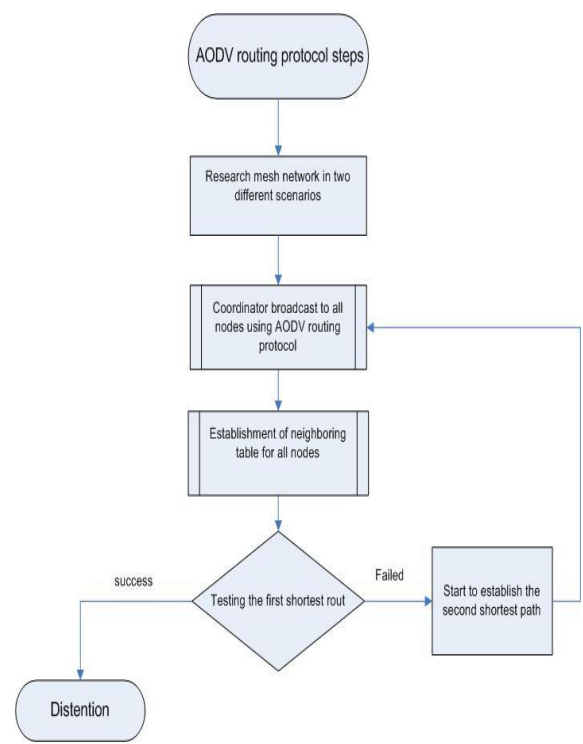

**Fig .1 Flow Chart of AODV Routing Schema.**

## 2. RELATED WORK

A considerable number of research works had been published on the effectiveness of mobile coordinator in WSNs, and most researchers concluded that mobile coordinator improves WSNs network performance [4, 6, 7, and 8]. None were concentrated on the impact of fixed positioning of the coordinator in ZigBee network which should be considered as an important step in improving the performance of ZigBee network.

Harsh Dhaka etal., performed an extensive evaluation study using OPNET, studying the impact of coordinator mobility in ZigBee mesh networks, their results indicate that the best performance had been achieved when the trajectory of the sink (coordinator) movement along fixed path, reflecting the fact that throughput was minimum for diagonals or square trajectories, also the best result achieved for lesser hops counts from the static sink [9].

Other researchers concentrated on the investigation of ZigBee network performance using static and random mobility coordinators in star, tree and mesh topologies. The authors concluded that random sink mobility will provide more performance than static sink because of neighboring nodes will consume their energy faster than other nodes located far away from the sink; also part of their conclusion was mesh topology was more efficient energy than star and tree [10].

For tree network topology, an experiment has been carried out to determine the effectiveness of positioning nodes among the network for the purpose of better throughput. The study was performed using OPNET simulator testing different mobility models, the authors concluded that group mobility model had the best performance among other suggested model such as random waypoint, random walk, and pursue mobility model [11]. From their literature, it had been concluded that various researchers have not taken into consideration the routing strategies and its optimization.

Jiasong Mu and Kaihua Liu in their publication research work was to simulate network performances by experimenting different route strategies such that Enable Route Discovery (ERD), Force Route Discovery (FRD) and Route Discovery (SRD). By means of changing node mobility and network dimensions using the evaluation tool of OPNET simulator, they concluded that ERD and AODV had the most efficiency and suitable for dynamic environments, FRD has always the worst performance, and SRD is the most suitable for static networks, and their suggestion was to concentrate on positioning status of nodes and coordinators mobility needs to be considered for the purpose of improving ZigBee network performance [12].

## 3. SIMULATION STEPS

In this research work MatLab software version R2013b was used to simulate AODV routing protocol in Mesh networks to study the routing process activities from different aspect of the problems assigned. Many factors had been taken in to consideration which played an important part in the assigned scenarios of the research work, the framework phases divided in three phases, I- Designing two networks and build neighboring table, II- Plotting node voltage decay and energy map III- Choosing two shortest routes and build a route connection.

### 3.1 Simulation Process

- Two different ZigBee Mesh network will be recognized, the first network will be created where the coordinator located at the center of the network and the second network coordinator located at the corner of the network for a span area of 600m x 600 m.
- The first Mesh network composed of one coordinator located at the center of the network shown in red color surrounded by 6 routers which represent first hop shown in green then surrounded by 8 end units shown in yellow , as shown in Figure.2
- The second network is the same as the first except that the coordinator will be located at left top corner of the network as shown in Figure.8
- Packets will be broadcast to all nodes and the first shortest route will be selected, if the first shortest fail due to battery strength below threshold value of 1.6 volts, a second scenario of second shortest route should be established.
- Plotting voltage battery decaying diagrams for scenarios, 1st and 2nd Shortest routes diagram for both networks.
- Plotting the energy map diagram, which represent battery strength of nodes in color for the weakest to the strongest.
- Building the neighboring table diagram which represents distances and voltage strength for all nodes.
- Distance calculation in meters for 1st, 2nd and other routes, used for comparison purposes.

### 3.2 Simulation Parameters

The network structure consists of network entities presented in Table 1.

**Table1. ZigBee Network Parameter**

| Parameter | Value |
|---|---|
| Network scale size | Fixed area (600×600 m). |
| Type of Technology | ZigBee / IEEE 802.15.4 |
| Network Topology | Mesh network |
| Routing Protocol used | AODV Routing Protocol |
| Operating Voltage Range | 1.65 -3.292 V |
| Power Supply | Power supply for Coordinator and Coin battery for end units and routers |
| Coin Battery Voltage | 3.292 Volts |
| Threshold | 1.6 Volts |
| Frequency | 2.4 GHz |
| Number of Bits per packet | 2000 Bits |
| Modulation type | On/ Off Keying Modulation scheme |
| Transmit power | 0.01w |
| Received power | 60 dbm |
| No. of Nodes | one coordinator, six routers and 8 end units |

## 4. SIMULATION ANALYSIS

The simulation results that illustrate the impact of using two different locations of the coordinator in ZigBee mesh networks, using AODV routing protocol, will be analyzed and results are detailed in the following sections.

## 4.1 The Coordinator at the Center of the Network

The first Mesh Network where the coordinator in red color located at the center of the network surrounded by 6 routers in green color, then surrounded by 8 end units in yellow color, the black line is the first shortest path, where the packets will be broadcasted to all nodes, until a route 1-2-8 will be established as shown in Figure .2.

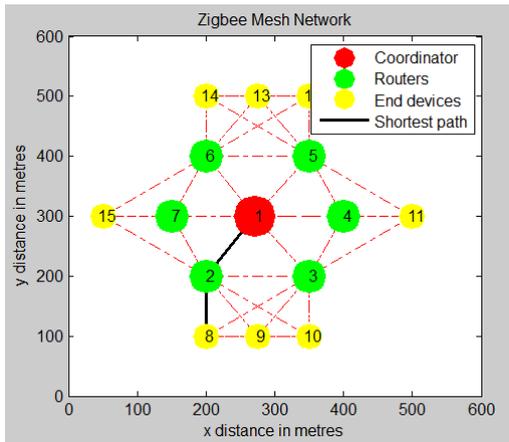

**Fig.2 First Mesh network the coordinator in the center with first shortest route.**

### 4.1.1 The Coordinator at the Center of the Network (version 1)

The coordinator will broadcast packets to all station for a total of 20000 transmission, each transmission time is assigned to be 0.02 seconds for a total transmission time of 400 seconds until a shortest path will be established, during all this process, the battery power for each nodes will be calculated using an optimized formula shown in Ref. [13] as shown in Figure.3.

As seen in the figure, node 1 is the coordinator where its voltage remain intact because it is powered by power supply, nod 2 is the first node of the shortest route, its voltage decayed more than other nodes, since it carry the shortest route, the destination which is node 8 will suffer less decay since it is an end unit doesn't work hard like routers. Routers 3, 4, 5, 6,7 will suffer more decay since it communicate with other router seeking best route since AODV routing protocol is used to provide discovery and maintenance process. Nodes 9, 10, 11, 12, 13, 14, 15 suffer less decay in battery voltage since they are end units operate less than routers.

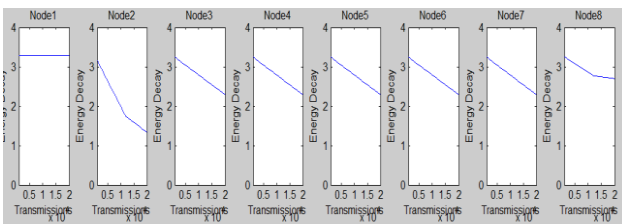

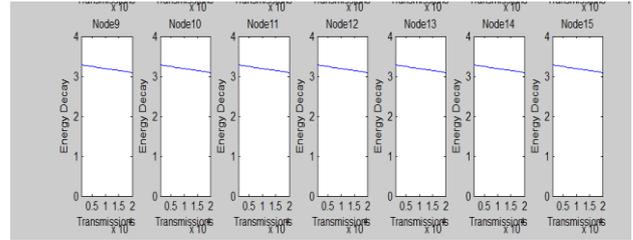

**Fig.3 Version 1: Voltage decaying for all nodes**

As shown in Table 2, the neighboring table will be established also, which is specifies the neighbor to each node, the first column represents the node number, the second column shows the neighbors and column 3 is the distance between the nodes and its neighbors measured in meters.

**Table 2: Snap shot of the Neighboring Table.**

| 'NODE' | 'NEIGHBOURS' | 'DISTANCES' |
|---|---|---|
| [ 1] | [ 2] | [ 122.0656] |
| [ 1] | [ 3] | [ 128.0625] |
| [ 1] | [ 4] | [ 130] |
| [ 1] | [ 5] | [ 128.0625] |
| [ 1] | [ 6] | [ 122.0656] |
| [ 1] | [ 7] | [ 120] |
| [ 2] | [ 1] | [ 122.0656] |
| [ 2] | [ 3] | [ 150] |
| [ 2] | [ 7] | [ 111.8034] |
| [ 2] | [ 8] | [ 100] |
| [ 2] | [ 9] | [ 125] |
| [ 2] | [ 10] | [ 180.2776] |
| [ 2] | [ 15] | [ 180.2776] |
| [ 3] | [ 1] | [ 128.0625] |
| [ 3] | [ 2] | [ 150] |
| [ 3] | [ 4] | [ 111.8034] |

Figure .4 represents the energy map for all nodes in the network; each color has its own battery percentage with respect of the maximum battery voltage (3.3 V), where the legend on the figure shows the percentages values, for example the pink color node has 60% left of its battery voltage which is calculated as 60% x 3.3 V= 1.98 V.

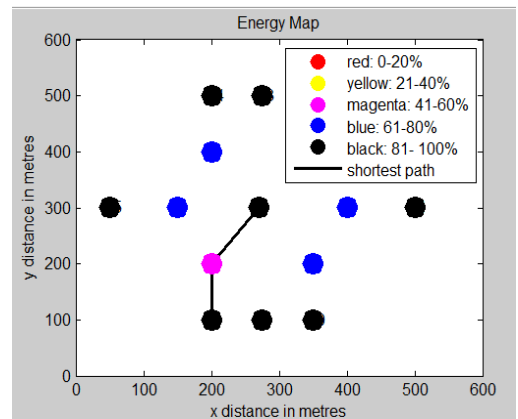

**Fig .4 Version 1: Network Energy Map for Coordinator in the center.**

## 4.1.2 The coordinator at the Center of the Network (Version 2)

During communication process of broadcasting packets to assigned destination and if the fist shortest route fails to reach its destination and since neighboring tables contains Information about distance and nodes voltages, a second shortest route will be established. Figure.5 shows both routes, the black line is the first shortest route, and the blue line is the second shortest route.

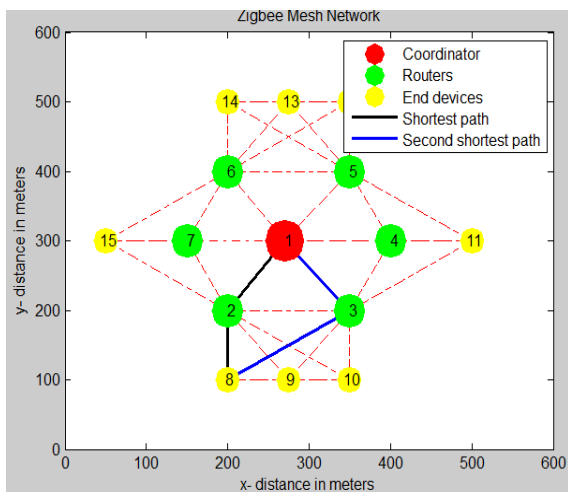

**Fig.5 Version 2: First and second shortest route.**

Table.3 show the distance in meters for the first, second and other routes, as seen the first route is the shortest.

**Table.3 1st, 2nd, and other r Routes Distances Calculated in Meter**

| Node | Distance |
|------|----------|
| 1-2 | 122.0656 |
| 2-8 | 100 |
| Sum | 222.0656 (First Route) |
| **Node** | **Distance** |
| 1-3 | 128.0625 |
| 3-8 | 180.2776 |
| sum | 308.3401 (Second Route) |
| **node** | **Distance** |
| 1-2 | 122.0656 |
| 2-3 | 150 |
| 3-8 | 180.2776 |
| Sum | 452.3432 (Other Routes) |

Figure.6 represents the voltage decaying of all nodes as a function of packet transmission for the Version 2. As seen node 3 voltage declines more due to its participation in routing process, node 8 voltage will decline little more since it has used for both routes.

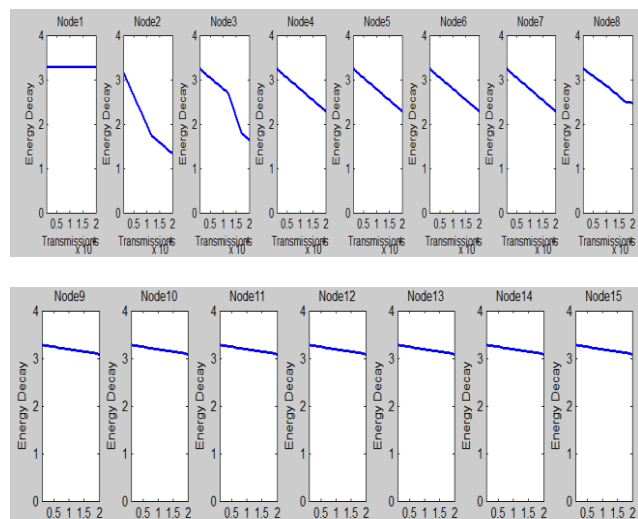

**Fig.6 Version 2: Voltage decaying of all nodes as a function of packet transmission.**

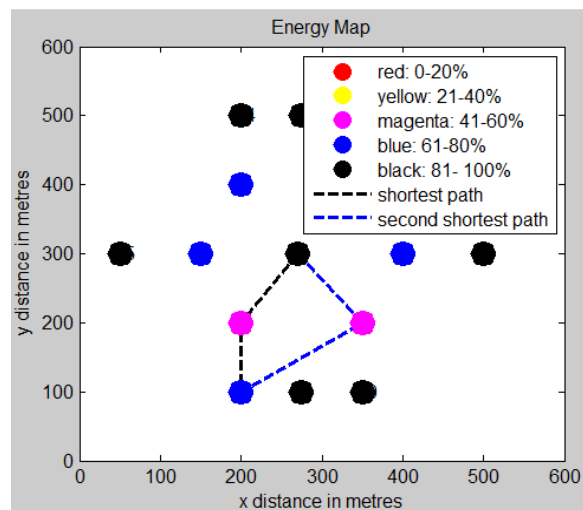

**Fig.7 Version 1: Network Energy Map for coordinator in the center.**

Figure.7 Energy map represented in percentage with respect to 3.3V of all nodes in colors as seen in the legend.

## 4.2. The Coordinator at the Corner of The Network

The same description and conditions as mentioned in section 4.1 will be applied to the second network where the coordinator is located at the corner of the network as shown in Figure. 8, the black line is the first shortest route which will go through nodes 1, 6, 7, 2 until it reaches its destination at node 8, as seen in figure the route is long and the packet will span through three different hops and the delay time will be getting larger.

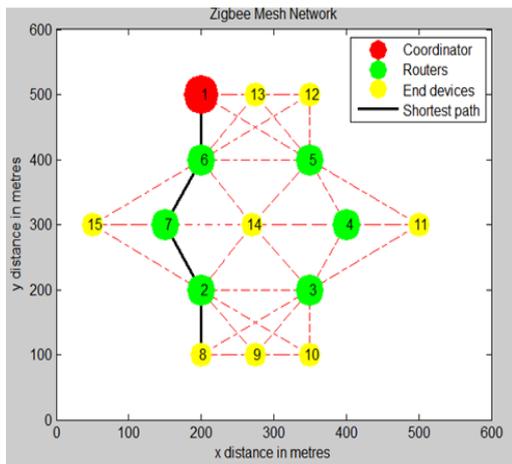

**Fig .8 Version 1 of the Second Mesh network where the coordinator in the corner.**

### 4.2.1 The Coordinator at the Corner of the Network (Version 1)

The voltage decline in battery power is shown in Figure.9, where batteries will consume some of its power during packets broadcast until first shortest route reaches its destination.

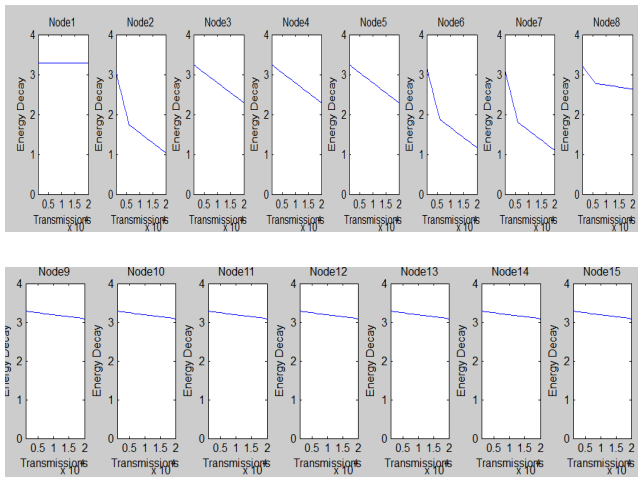

**Fig.9 version 1: Voltage decaying of nodes batteries for the first network where the coordinator located at the center of the network**

### 4.2.2 The coordinator at the corner of the Network (Version 2)

Version 2 is the case where the first shortest route fails to reach its destination due to battery voltage decline, a second shortest route will be established to reach its destination, during all this process a neighboring table as shown in Table.4 will be created which contain all distance between nodes and battery voltage, Figure.10 shows both routes, the black line is the 1st shortest route, and the blue line is the second shortest route, as seen the second route is longer and both routes spanning through 3 hops and its delay time in longer, Table .5 represent the 1st, 2nd and other route distances.

**Table .6 Snap Shut from the neighboring Table in version 2**

| 'NODE' | 'NEIGHBOURS' | 'DISTANCES' | 'ENERGY' |
|---|---|---|---|
| [ 1] | [ 2] | [ 122.0656] | [3.2920] |
| [ 1] | [ 3] | [ 128.0625] | [3.2920] |
| [ 1] | [ 4] | [ 130] | [3.2920] |
| [ 1] | [ 5] | [ 128.0625] | [3.2920] |
| [ 1] | [ 6] | [ 122.0656] | [3.2920] |
| [ 1] | [ 7] | [ 120] | [3.2920] |
| [ 2] | [ 1] | [ 122.0656] | [1.3383] |
| [ 2] | [ 3] | [ 150] | [1.3383] |
| [ 2] | [ 7] | [ 111.8034] | [1.3383] |
| [ 2] | [ 8] | [ 100] | [1.3383] |
| [ 2] | [ 9] | [ 125] | [1.3383] |
| [ 2] | [ 10] | [ 180.2776] | [1.3383] |
| [ 2] | [ 15] | [ 180.2776] | [1.3383] |
| [ 3] | [ 1] | [ 128.0625] | [1.6442] |
| [ 3] | [ 2] | [ 150] | [1.6442] |
| [ 3] | [ 4] | [ 111.8034] | [1.6442] |
| [ 3] | [ 8] | [ 180.2776] | [1.6442] |

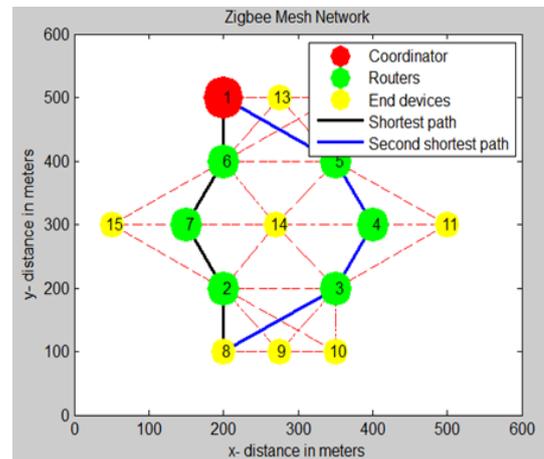

**Fig.10 Versions two of the Second Mesh network.**

Figure .10 shows the 1st shortest route in black and the second shortest route in blue color, the coordinator is shown in red color located at the corner of the network linked with 6 green colored routers and 8 yellow colored end units.

**Table.5 1st, 2nd and other Routes distance calculations in meters.**

| Node | Distance |
|---|---|
| 1-6 | 100 |
| 6-7 | 111.8034 |
| 7-2 | 111.8034 |
| 2-8 | 100 |
| Sum | 423.6068 (First Route) |
| **Node** | **Distance** |
| 5-1 | 180.2776 |
| 5-4 | 111.8034 |
| 4-3 | 111.8034 |

| 3-8 | 180.2776 |
| Sum | 584.16 (second route) |
| **Node** | **distance** |
| 1-6 | 100 |
| 6-5 | 150 |
| 5-4 | 111.8034 |
| 4-3 | 111.8034 |
| 3-8 | 180.2776 |
| Sum | 473.081 (Other Routes) |

Figure.11 shows the decline in voltage value during the second route discovery, as seen the nodes voltage decline even further due to its participation in 1$^{st}$ and 2$^{nd}$ route process.

Figure.12 shows the energy maps for all nodes after the 2$^{nd}$ shortest path takes place, the legend show the percentage of remaining voltages in different colors.

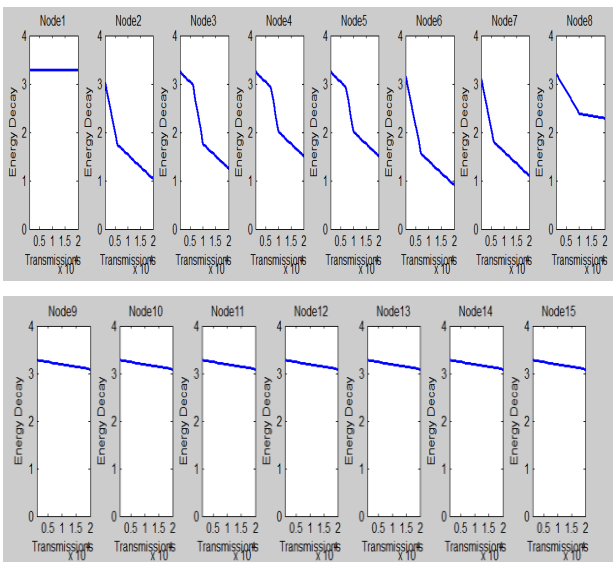

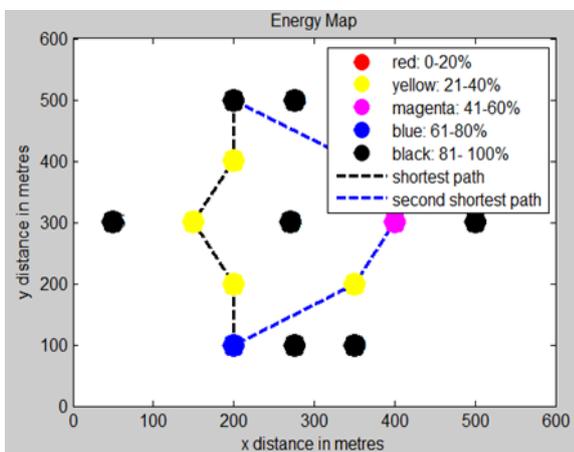

**Figure .11 versions 2: voltage decaying of nodes batteries for the second network where the coordinator located at the corner of the network**

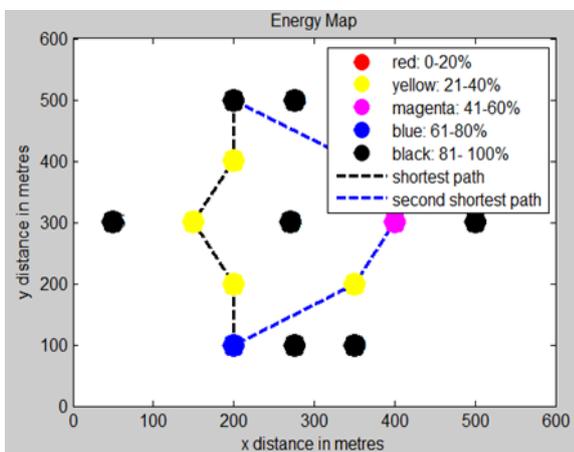

**Fig.12 Energy Map represented in percentage for all nodes after the 2$^{nd}$ shortest path.**

## 5. CONCLUSION

The most important factor in of AODV routing in ZigBee Mesh network depends mainly on network lifetime which depends mainly on node batteries and proper network design, when the coordinator initiate the network operation and broadcast packets, the routes need some time until it reaches its destination or may be the route could not get established due to improper Mesh Topology design or node batteries voltage decaying.

In this research work the factors that could increase network longevity and improve network performance were studied. Two network been built for two different coordinator location, our results concluded that the central location of the coordinator within the network is more appropriate for reliable communications between nodes and the shortest route could be generated fast which lead to fast communication and less power consumption in nodes and more nodes could retain its energy which will lead to a longer life for the network, the established neighboring table during routing process of AODV protocol could support the selection of the nearest nodes that have enough power and shortest distance to the transmitting node in order to build a route connection between source and destination also, the energy mapping has great benefits to recognize weak nodes from strongest one, since the energy map shows the percentage of voltages left in nodes batteries.

## 6. REFERENCES

[1] Yahaya, F. H, Yussoff, Y .M; Rahman RA, and Abidin N.H.2014.Performance Analysis of Wireless Sensor Network. Signal Processing and its Applications IEEE; March 400-405; Kuala Lumpur, Malaysia.

[2] Daintree.2015.Getting Started with ZigBee and IEEE 802.15.4.http://www.daintree.net/downloads/whitepapers /ZigBee_primer.pdf, Accessed March.

[3] William C. 2008.Wireless Control That Simply Works. White paper, ZMD America Inc.

[4] Liang NC, Chen PC, Sun T, Yang G, Chen LJ, Gerla M. 2006.Impact of node heterogeneity in ZigBee mesh network routing. IEEE International Conference on Systems Man and Cybernetics; pp.187-191; Taipei.

[5] Kinney P. 2003.ZigBee Technology: Wireless Control that Simply Works. Communications Design Conference; pp.1-20; Calif, USA.

[6] Luo J, Panchard J, Piorkowski M, Grossglauser M, Hubaux JP. 2006. Mobiroute: Routing Towards a Mobile Sink for Improving Lifetime in Sensor Networks. 5th IEEE International Conference In Distributed Computing in Sensor Systems Springer.pp. 480-497; USA.

[7] Shakya. M, Zhang. J, Zhang. P, Lampe. M. 2007.Design and optimization of wireless sensor network with mobile gateway. IEEE 21$^{st}$ International Conference in Advanced Information Networking and Applications Workshops; pp.415-420; Niagara Falls, Canada.

[8] Bi. Y, Sun. L, Ma. J, Li N, Khan. I, Chen. C. 2007.HUMS: an autonomous moving strategy for mobile sinks in data-gathering sensor networks. EURASIP Journal on Wireless Communications and Networking, vol. 1, pp. 064574.

[9] Dhaka. H, Jain. A, Verma. K., 2010. Impact of Coordinator Mobility on the Throughput in a ZigBee Mesh networks. IEEE Conference in Advance Computing Conference, pp. 279-284, Patiala.

[10] Aziz. A, Qureshi. M. A, Soorage M.U, Kashif. M.N, Hafeez MA., 2012.Evaluation of ZigBee Based Wireless Sensor Network with Static Sink and Random Sink Mobility. International Journal of Computer and Electrical Engineering.; Vol.4, no.4, pp.562-566.

[11] Parneet D, Sadawarti H. 2014.Impact Analysis on the Performance of ZigBee Protocol under Various Mobility Models, International Journal of Engineering Trends and Technology (IJETT).vol. 9, No. 11, pp.550- 562.

[12] Mu .J, Liu. K. 2010. Effect of node mobility and network dimension to the Zigbee Routing Method. 6th International Conference on. IEEE in Wireless Communications Networking and Mobile Computing (WiCOM). pp.1-5; Chengdu.

[13] Hussein. A, Samara. G. 2015.Mathematical Modeling and Analysis of ZigBee Node Battery Characteristics and Operation, MAGNT Research Report.; vol.3, No.6, pp. 99-106.